# General and scalable vapor etching and transformation platform for two-dimensional materials

Zhiguo Du[1]†, Jikai Zhang[1]†, Jonas Björk[2,3], Zongju Cheng[1], Ningjun Chen[2], Qi Zhao[1], Hao Chen[1], Yuxuan Ye[1], Guang Yang[4], Haiyang Wang[1], Bin Li[1], Johanna Rosen[2,3]* and Shubin Yang[1]*

[1]School of Materials Science and Engineering, Beihang University; Beijing 100191, China.

[2]Materials Design Division, Department of Physics, Chemistry and Biology (IFM), Linköping University; Linköping 58183, Sweden.

[3]Wallenberg Initiative Materials Science for Sustainability (WISE), Department of Physics, Chemistry and Biology (IFM), Linköping University; Linköping 58183, Sweden.

[4]JiNan SCMXene Tech. Co., Ltd.; Jinan 250307, China.

*Corresponding author. Email: johanna.rosen@liu.se, yangshubin@buaa.edu.cn

†These authors contributed equally to this work

**Abstract:** Two-dimensional (2D) nanomaterials derived from non-van der Waals (non-vdW) solids offer exceptional physicochemical properties, yet their synthesis is impeded by intrinsic covalent/metallic bonding and high surface reactivity of the precursors. Here, we report a general vapor-phase etching and transformation platform for producing a library of 36 2D carbides, nitrides, and carbonitrides, exhibiting electrical conductivities spanning six orders of magnitude. Using reactive vapors like hydrogen chloride, we selectively remove A-layers from MAX phases to yield well-defined layers (MXenes), including previously inaccessible semiconducting $Hf_2CT_x$. By varying the reactive vapor environment, MXenes can be engineered at X-site and surface-termination site and even be transformed into non-vdW layers such as 2D MAX phases. This general and scalable vapor-phase platform reframes 2D material synthesis, opening new avenues for various applications.

Two-dimensional (2D) nanomaterials originating from van der Waals (vdW) and non-vdW solids have attracted considerable interest owing to their unique physical and chemical properties that deviate drastically from their bulk counterparts (*1–5*). Atomic layers from vdW solids, such as graphene and transition metal dichalcogenides (TMDs), are distinguished by the absence of dangling bonds on the surfaces, and can be readily exfoliated from their layered vdW counterparts (*6*, *7*). In contrast, 2D materials from non-vdW laminates are typically more challenging to obtain due to strong out-of-plane chemical bonds (e.g., covalent or metallic bonding) (*8*, *9*). In the process, their synthesis entails cleavage of these intense interactions, resulting in highly reactive surfaces with unsaturated coordination and abundant dangling bonds, necessitating the termination of these layers with surface functional groups (*10*, *11*). Notably, the atomic structure and surface terminations of 2D materials from non-vdW solids can be tailored through precursor design or post treatment, making them highly promising for catalysis, energy storage, electromagnetic interference shielding and superconductivity (*12–15*).

Atomic layers not originating from vdW compounds, such as 2D monoelemental layers (e.g., silicene and bismuthene) (*16*, *17*), metal oxides (e.g., magnetene and hematene) (*18*, *19*), and group III–V compounds (e.g., 2D GaN) (*20*, *21*), have been produced through exfoliation (*22*, *23*), physical/chemical vapor deposition (*24*, *25*) and vdW squeezing methods (*17*, *20*). As a representative class of such atomic layers, 2D transition-metal carbides, nitrides, and carbonitrides (MXenes) were initially synthesized via a wet chemical etching approach, which involves the selective removal of A-layers from non-vdW MAX phases (where M is a transition metal, A is an A-group element and X is C and/or N) in a liquid medium, enabling the synthesis of O-, OH-, and F-terminated MXenes (*26–28*). Later, a molten salt etching approach was explored to selectively etch MAX phases based on the redox potential coupling of A species with Lewis acidic salts ($ZnCl_2$, $CuCl_2$ or $CdBr_2$), allowing the production of Cl- and Br-terminated MXenes (*29–31*). Moreover, such F-free MXenes can be further functionalized with -S, -Se or -Te groups, leading to tailored electrical properties such as superconductivity (*32*). More recently, a direct synthesis route has been developed for chemical vapor growth of MXene by the reaction of metals and metal halides with carbon/nitrogen sources (e.g., graphite, methane, or nitrogen), circumventing the conventional reliance of MAX phases in MXene production (*33*). Although the above three approaches can be used to produce some MXenes, they still have many intrinsic limitations such as tedious washing or high cost. Thus, it remains highly desirable to develop new synthesis protocols for the realization of MXenes, and beyond, through fast and scalable procedures.

## Synthesis of MXenes by vapor etching

To address the limitations of the known synthetic methods of MXenes, we develop a vapor etching approach to produce MXenes under hydrogen chloride (HCl) gas (see Safety Measures in Supplementary Materials). At elevated temperatures (> 873 K), HCl gas rapidly reacts with the A element in MAX phases, forming volatile A-containing chlorides and releasing hydrogen, while preserving the layered M-X framework to yield MXene (MAX + HCl (g) → MX (MXene) + $ACl_y$ (g) + $H_2$ (g); Fig. 1A and fig. S1). The formation of volatile $ACl_y$ species during the reaction facilitates their removal from the solid matrix, preventing byproduct accumulation and enabling deep HCl penetration, which in turn accelerate the kinetics to achieve sustained and uniform etching (fig. S2). To understand the thermodynamic feasibility of this vapor etching protocol, we performed density functional theory (DFT) calculations using $Ti_4AlN_3$ as a representative model (Fig. 1C). A thermodynamic comparison of Gibbs free energy changes ($\Delta G$) was performed for two competing pathways: the selective etching of A elements to form MXene versus the total disintegration of the pristine MAX phase. The results demonstrate that $\Delta G$ for the transformation from MAX phase to MXene (MAX→MXene) remains consistently lower than that of the disintegration pathway across all calculated HCl-to-MAX molar ratios ($n_{HCl}/n_{MAX}$). Notably, as increasing the HCl concentration, the thermodynamic advantage of the MAX→MXene process becomes more distinct, confirming the energetic preference for MXene formation under our reaction conditions. Such kinetic and thermodynamic advantages underpin the potential scalability of the process, as demonstrated by a single-batch synthesis of 285 grams in our lab (fig. S1). We also compared the vapor etching approach with representative MXene synthesis methods (*26,27,29,30,33*) – including wet chemical etching, molten salt etching, and chemical vapor deposition (CVD) – using a radar chart focused on sustainability metrics (Fig. 1D). This assessment highlights that our vapor etching route exhibits many advantages across multiple dimensions, particularly in cost-effectiveness, water reduction, environmental friendliness, and process simplicity, while maintaining high versatility in the types of MXenes produced, as exemplified by the diverse library discussed later.

As a proof of concept, $Ti_4AlN_3$ was reacted with HCl gas at 923 K for 25 min to produce nitride MXene (Fig 2A; see Materials and Methods in Supplementary Materials). Scanning electron microscopy (SEM) (Fig. 2B and fig. S3) and transmission electron microscopy (TEM) images (fig. S4) show an accordion-like morphology, reminiscent of MXenes obtained via a wet chemical etching approach (*26*). This delaminated structure likely originates from the rapid removal of volatile aluminum chloride and hydrogen gas during the etching process (fig. S5). X-ray diffraction (XRD) patterns of the product exhibit a broad diffraction peak at 4.7° (Fig. 2C), corresponding to the (002) peak of MXene, shifted from 7.5° in $Ti_4AlN_3$, due to interlayer expansion after Al removal. Remarkably, the etching process occurs within only 25 mins under reactive vapor, substantially faster than the 18–48 h and 5–24 h required for wet chemical and molten salt etching approaches, respectively (*26*, *27*, *30*).

Owing to the weakened interlayer interactions, the as-prepared MXene can be readily delaminated into 2D atomic layers with a yield of ~24.5 wt.% via a simple sonication (Fig. 2, D and E, and fig. S6). Atomic force microscopy (AFM) image and corresponding height profiles (Fig. 2D) present thin layers with average thicknesses of 1.4–1.5 nm, corresponding to MXene monolayers. High-resolution TEM (HRTEM) images show well-defined lattice fringes with a spacing of 0.27 nm in the delaminated layers (fig. S6F), matching the lattice spacing of the (100) planes. Atomic-resolution scanning transmission electron microscopy (STEM) image shows the transition metal atoms arranged in a hexagonal symmetry (Fig. 2F and fig. S7), disclosing a single-crystalline feature (inset in Fig. 2F). Elemental mapping images demonstrates uniform distribution of Ti, N and Cl (fig. S6, H to J), consistent with X-ray absorption spectroscopy (XAS) and X-ray photoelectron spectroscopy (XPS) analyses (figs. S8 and S36), and elemental analysis indicates a Cl content of 10.4 wt.%. These results confirm that the MXene is primarily terminated with Cl species, possessing a Ti-N-Cl configuration. In addition, O terminations are present on the nitride MXene (fig. S6K), arising from trace $H_2O$ impurities in the HCl gas. Accordingly, we designate the resultant MXene as $Ti_4N_3T_x$ ($T_x$ = Cl, O).

## MXene layers featuring M-X-Cl configuration

To understand the generality of the vapor etching platform, we developed a theoretical framework combining density functional theory (DFT) calculations of MAX and MXene free energies with thermochemical data for gas-phase species (see details in Supplementary Materials). The model compares two competing reactions – formation of MXenes versus complete disintegration to volatile chlorides – as a function of HCl pressure, temperature and the initial HCl-to-MAX molar ratio $n_{HCl}/n_{MAX}$. The calculated results are summarized in a thermodynamic stability map (Fig. 3A, and figs. S9 to S35), where the vertical and horizontal axes represent the $\Delta G$ for MXene formation and MAX disintegration, respectively. In this representation, the diagonal serves as the thermodynamic boundary. Data points situated below the diagonal signify that MXene formation is energetically more favorable than disintegration, while those above indicate a preference for the latter. Consistent with these criteria, we experimentally synthesized 12 MXenes with distinct atomic structures, including Ti-, Nb-, Ta-, Zr- and Hf-based carbides, carbonitrides and nitrides (figs. S36 to S51, and table S1). In contrast, MAX phases positioned above the diagonal, such as $V_2AlC$ and $Cr_2AlC$, failed to yield MXenes due to the thermodynamic preference for disintegration (fig. S52). In addition, MAX phases containing Mo exhibit low reactivity towards both exfoliation and disintegration, likely due to the relatively weak bonding of Cl to Mo-based MXenes compared to other MXenes (*34*). Typically, atomic-resolution STEM images of $Ti_4N_3T_x$, $Ti_3C_2T_x$ and $Zr_2CT_x$ show the well-defined layered architectures (Fig. 3, B to D, figs. S37, S40, and S50), in which the bright transition metal atoms are symmetrically flanked by surface terminations. The interlayer spacings vary in the range from 1.31 to 1.91 nm for $Ti_4N_3T_x$, and those of $Ti_3C_2T_x$ and $Zr_2CT_x$ are 1.14 and 0.98 nm, respectively.

Notably, Zr and Hf-based MXenes, such as $Zr_2CT_x$ and $Hf_2CT_x$, have garnered significant theoretical attention owing to their predicted unique semiconducting properties (*35*, *36*). However, they have remained experimentally elusive via conventional methods due to the limited etching selectivity between the transition metal and A species (*37*). Our vapor etching strategy overcomes this challenge and enables the synthesis of the above two types of MXenes (see Methods). To the best of our knowledge, this is the first experimental synthesis of $Hf_2CT_x$ MXene. The as-produced 12 MXenes reveal a wide-range variation in electrical conductivities, spanning six orders of magnitude (Fig. 3E). Thereinto, 10 MXenes exhibit conductivities on the order of $10^3$ to $10^4$ S $m^{-1}$, featuring a metallic character. In stark contrast, the $Zr_2CT_x$ and $Hf_2CT_x$ show a drastic drop to the $10^{-2}$ S $m^{-1}$ magnitude, signaling a definitive transition into the semiconducting regime. To further verify the semiconducting nature, we performed UV-vis diffuse reflectance spectroscopy measurement to evaluate the optical bandgap. The derived Tauc plots (Fig. 3, F and G) validate bandgaps of 2.04 and 2.24 eV for $Zr_2CT_x$ and $Hf_2CT_x$, respectively. These results confirm their identity as indirect bandgap semiconductors, expanding the functional scope of MXene family beyond its traditional metallic members.

**Tailored 2D layers with M-X-T′ (T′=O, S, Se, Te, P) configurations**

Within the framework of our vapor-phase route, we further achieve post-synthetic tailoring of the surface termination compositions in MXenes by introducing reactive gases or vapors such as $O_2$, $H_2S$, Se, Te and P into the reaction system (**Fig. 4**a). As a representative example, $Ti_4N_3T_x$ MXene, upon exposure to $O_2$ at 523 K, undergoes a configuration transformation from M-X-Cl to M-X-O. X-ray absorption analyses confirm the replacement of Ti-Cl with Ti-O bonds (Fig. 4B and figs. S53 to S56), and elemental analysis shows a significant increase in oxygen level from 3.7 to 14.0 wt% after $O_2$ exposure. Such transformation is consistent with theoretical predictions, which indicate a thermodynamic preference for substituting -Cl with -O species under an oxygen atmosphere (*34*). This alteration endows the MXene with a Ti-N-O configuration that exhibits semiconductor-like behavior (figs. S57 to S61), distinctly different from the Ti-N-Cl configuration with metallic conductivity. Importantly, unlike previously reported labile atomic layers, MXene with a Ti-N-O configuration exhibit excellent chemical and thermal stability, tolerating basic solutions for up to three weeks (figs. S62 and S63) and remaining intact at temperatures up to 1100 K (fig. S64). Based on the available gases or vapors mentioned above, the transformation protocol enables the synthesis of a broad range of 2D atomic layers, including 11 MXenes with M-X-S, M-X-Se, M-X-Te, and M-X-P configurations (figs. S65 to S68), as well as non-vdW atomic layers with Ti-N and Ti-C rocksalt-type structures (figs. S69 and S70).

**X-site-substituted 2D layers and M-X to M-A-X conversion**

Beyond surface engineering, the vapor transformation strategy enables modification of the internal X-site species through atomic substitution. By introducing $CH_4$ or $NH_3$ into the reaction system, we implant C or N species into nitride or carbide MXenes, respectively. High-resolution N 1s XPS spectra clearly reveal the emergence of characteristic peaks corresponding to N-Ti bonds in MXenes with Ti-C-Cl configurations, such as $Ti_3C_2T_x$ and $Ti_2CT_x$ (Fig. 4C, figs. S71D and S71F). Moreover, high-resolution C 1s XPS spectra disclose the C-Ti peaks in Ti-N-Cl MXenes (fig. S71B). These results demonstrate the incorporation of nitrogen and carbon species into carbide and nitride MXenes, respectively. Such atomic exchange provides a route for interconversion between carbide and nitride frameworks. Furthermore, the reactivity of MXenes with an M-X-Cl configuration toward metal vapors, such as Sn, was explored. After reaction at 1073 K, $Ti_3C_2T_x$ exhibits the disappearance of the (002) peak at 8.0° in the XRD patterns (fig. S72), while the derived product shows the emergence a new weak, broad peak at 9.4°, corresponding to the (002) reflection of the $Ti_3SnC_2$ MAX phase (*38*). Atomic-resolution STEM images (Fig. 4, D to G) reveal that a single Sn atomic layer occupies the space between adjacent MX slabs, producing non-vdW atomic layers resembling ultrathin MAX slabs (MAXenes), with the general cross-sectional formula of Cl-MX-[A-MX]$_n$-Cl, where the integer $n$ indicates the number of inserted A layers. From a cross-sectional perspective, we observe MAXene structures containing a single A layer (Cl-MX-A-MX-Cl, Fig. 4d), two A layers (Cl-MX-A-MX-A-MX-Cl, Fig. 4E), and even four A layers (Cl-MX-A-MX-A-MX-A-MX-Cl, figs. S73), with the precise stacking dictated by the underlying crystalline symmetry of the MX slabs. Notably, the outer surfaces remain -Cl terminated rather than being substituted by Sn (fig. S73), indicating that Sn serves primarily as a zipper between adjacent MX slabs with low steric hindrance (fig. S74), rather than substituting the -Cl terminations. This process can be regarded as a form of atomic-level additive manufacturing between layers, enabling synthesis of a variety of non-vdW atomic layers with precise structures, such as 2D $Ti_3AlC_2$ layers produced via Al vapor (figs. S75 to S78). The as-produced MAXenes exhibit high electrical conductivities of up to 47600 S $m^{-1}$, exceeding those of the commercial $Ti_3AlC_2$ powder (4390 S $m^{-1}$) as well as MXenes with M-X-Cl (29200 S $m^{-1}$) and M-X-O (160 S $m^{-1}$) configurations.

**Scalable production**

Producing MXenes and other 2D layers from non-vdW solids often requires tedious washing steps to remove by-products and residues. Although feasible at the laboratory scale, such procedures pose a significant barrier to industrial-scale production. In contrast, we establish vapor etching and transformation as a general, solvent-free, and scalable platform for converting non-van der Waals layered solids into a diverse library of tailored two-dimensional carbides, nitrides, and carbonitrides, enabling routine production of up to 10 kilograms in our laboratory (Fig. 1B). The resulting materials span metallic to semiconducting behavior over six orders of magnitude in conductivity, including previously inaccessible members such as $Hf_2CT_x$. Beyond expanding the MXene family, this work reframes the synthesis of 2D materials, with a simple, washing-free approach that could be decisive for the commercial scale production of 2D atomic layers at low cost. Looking ahead, integration of vapor-phase routes with the continued development of non-vdW solids such as MAX phases and other laminates, this strategy is expected to substantially expand applications of MXenes and other 2D layers across catalysis, energy storage, electromagnetic interference shielding, interfacial engineering, and microwave absorption (fig. S79).

**Acknowledgments:** We thank the support from Beijing Synchrotron Radiation Facility and National Synchrotron Radiation Laboratory (NSRL, Hefei, China). We are grateful to the Analysis & Testing Center of Beihang University for the facilities, and the scientific and technical assistance.

**Funding:** This work was supported by National Natural Science Foundation of China grant 52125207 (S.Y.); National Natural Science Foundation of China grant 22475009 (Z.D.), Beijing Natural Science Foundation grant 2242039 (Z.D.); European Union (ERC, MULTI2D, 101087713); Knut and Alice Wallenberg (KAW) Foundation (support of the Wallenberg Initiative Materials Science for Sustainability); and the Swedish Research Council 2022-06725 (computational resources through the National Academic Infrastructure for Supercomputing in Sweden).

**Author contributions:**

Conceptualization: S.Y., J.R.

Methodology: Z.D., J.Z., J.B., S.Y., J.R.

Investigation: Z.D., J.Z., J.B., Z.C., N.C., Q.Z., H.C., Y.Y., G.Y., H.W., B.L.

Visualization: Z.D., J.Z., J.B.

Funding acquisition: S.Y., J.R., Z.D.

Project administration: S.Y., J.R., Z.D., J.Z., N.C.

Supervision: S.Y., J.R.

Writing – original draft: Z.D., J.Z., J.B.

Writing – review & editing: S.Y., J.R., Z.D., J.Z., J.B.

**Competing interests:** Authors declare that they have no competing interests.

**Data, code, and materials availability:** All data are available in the main text or the supplementary materials.

## Supplementary Materials

Materials and Methods

Figs. S1 to S79

Tables S1 to S3

References (*39–63*)

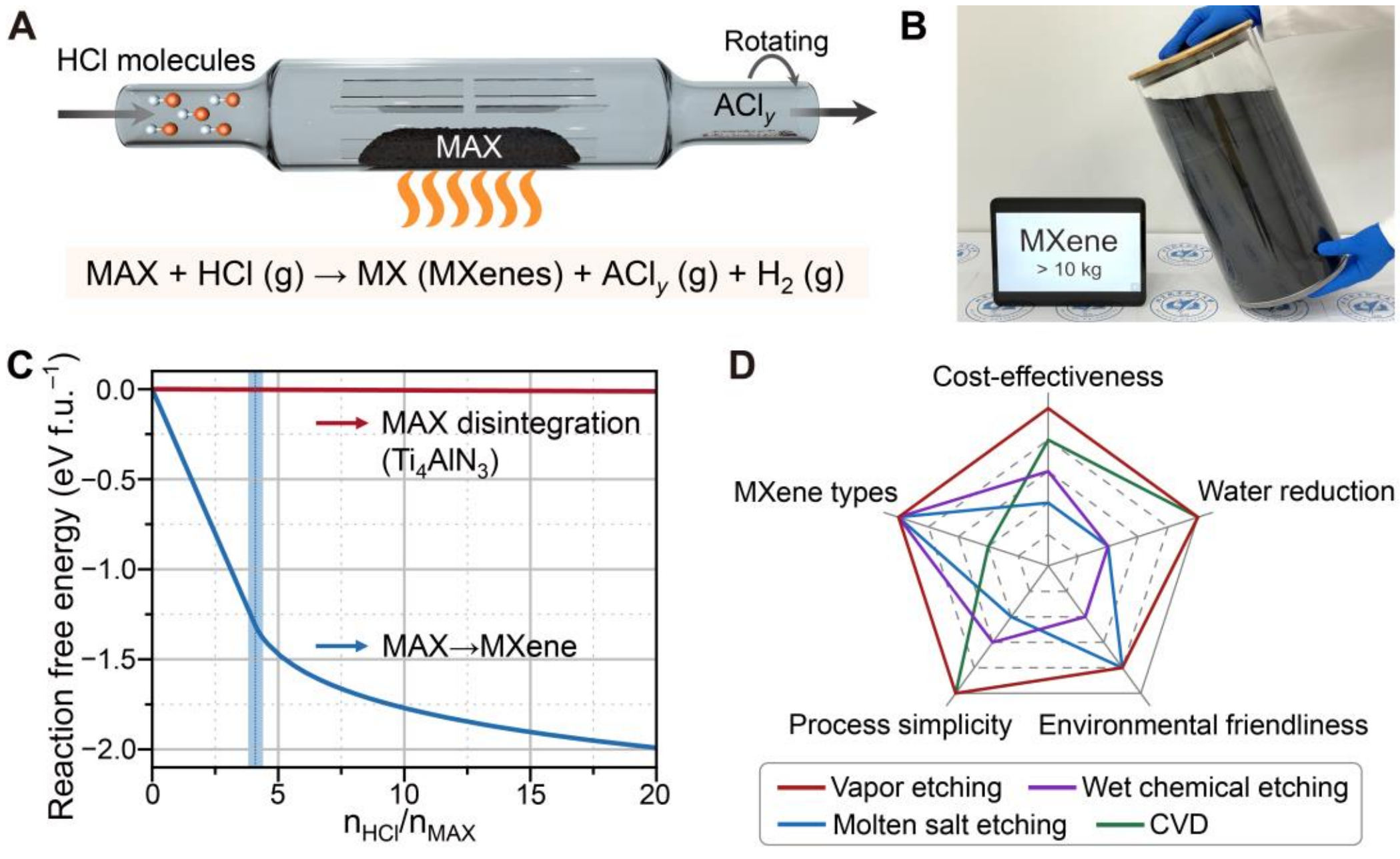

**Fig. 1. Vapor etching platform to produce MXenes under hydrogen chloride gas.** (**A**) MAX phases are progressively transformed to MXenes under HCl gas via the reaction (MAX + HCl (g) → MX (MXene) + $ACl_y$ (g) + $H_2$ (g)) in the temperature range of 873–1073 K, associated with volatile by-products of $ACl_y$. (**B**) Photograph of > 10 kilograms of MXene produced through the vapor etching approach at 285 g in one batch. (**C**) Reaction free energy of $Ti_4AlN_3$ under HCl gas, showing more negative Gibbs free energy changes for MXene formation (MAX→MXene) than disintegration (MAX dis.) as a function of HCl-to-MAX molar ($n_{HCl}/n_{MAX}$). (**D**) Radar chart, comparing the vapor etching method with representative MXene synthesis techniques (wet chemical etching, molten salt etching and CVD) across five dimensions: cost-effectiveness, water reduction, environmental friendliness, process simplicity and MXene types.

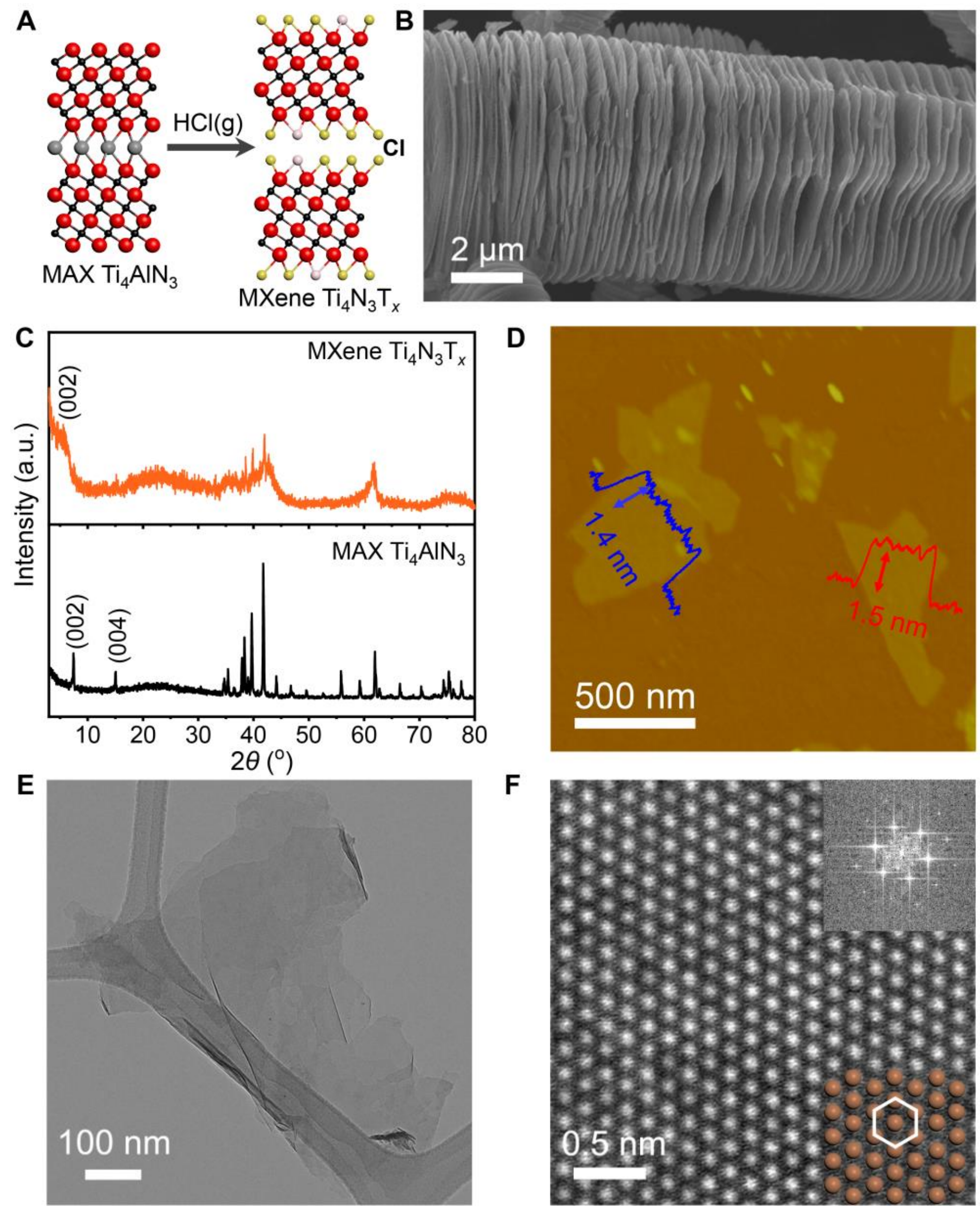

**Fig. 2. Characterizations of a typical MXene $Ti_4N_3T_x$ synthesized by vapor etching of $Ti_4AlN_3$.** (**A**) Schematic illustration of the transformation of MAX $Ti_4AlN_3$ to MXene $Ti_4N_3T_x$ under HCl gas. (**B**) SEM image of $Ti_4N_3T_x$, showing a typical accordion-like structure. (**C**) XRD patterns of MAX $Ti_4AlN_3$ and accordion-like $Ti_4N_3T_x$, exhibiting the presence of typical (002) peak for MXene. (**D**) AFM image and corresponding height profiles of the delaminated MXene $Ti_4N_3T_x$, exhibiting monolayers with average thicknesses of 1.4 and 1.5 nm. (**E** and **F**) TEM (E) and atomic-resolution STEM (F) images of MXene $Ti_4N_3T_x$, disclosing the precise atomic structure in a hexagonal symmetry. Insets in (F) are the corresponding FFT patterns (top) and atomic configuration (bottom).

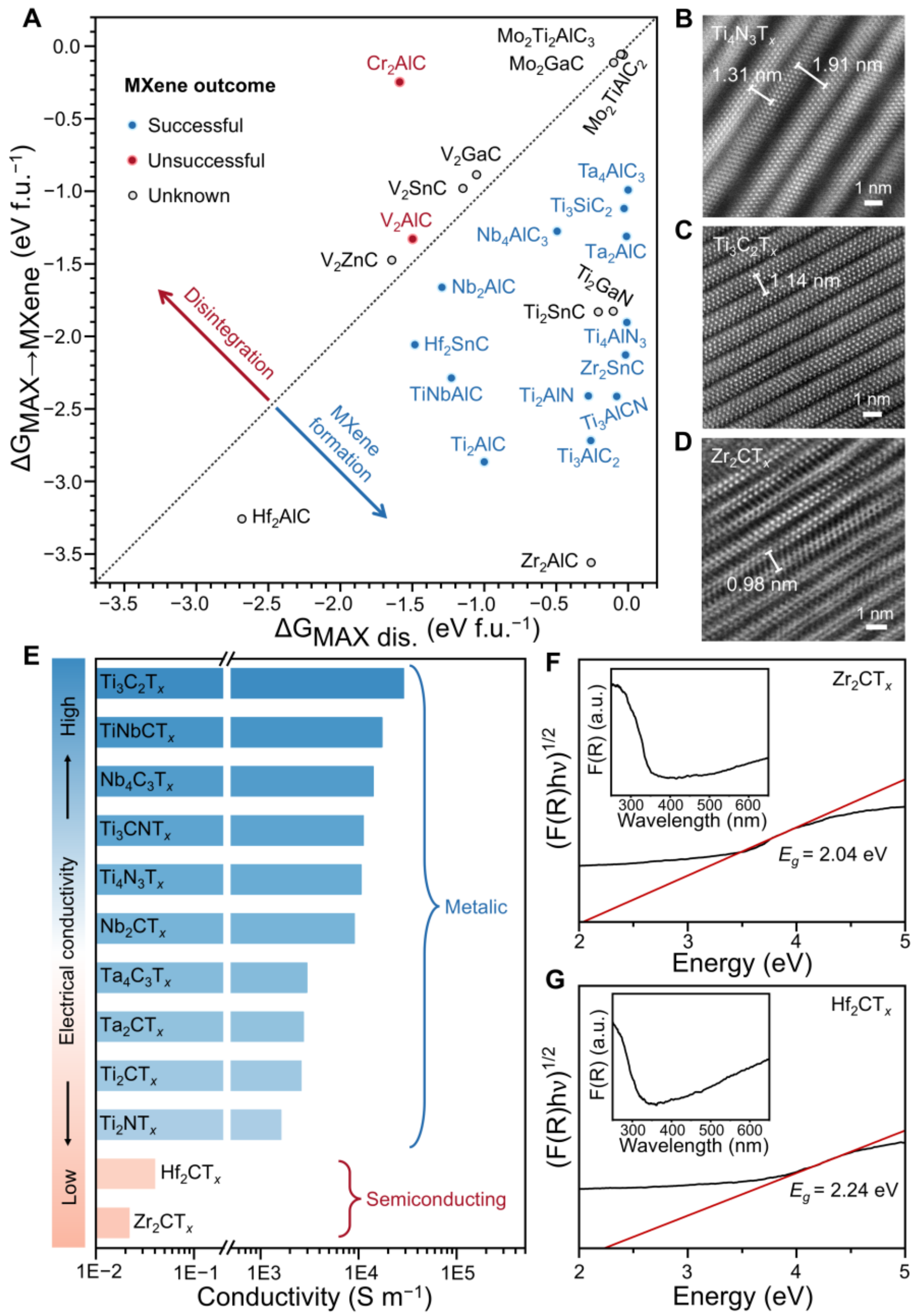

**Fig. 3. Universal synthesis of MXenes and discovery of semiconducting members.** (**A**) Thermodynamic stability map for the etching process of MAX phases under HCl gas, in which the axes represent calculated $\Delta G$ for MXene formation and MAX phase disintegration. (**B** to **D**) Atomic-resolution cross-sectional STEM image of $Ti_4N_3T_x$ (B), $Ti_3C_2T_x$ (C), and $Zr_2CT_x$ (D), showing typical layered structure with expanded interlayer spacings. (**E**) Electrical conductivities of 12 as-prepared MXenes, spanning six orders of magnitude. (**F** and **G**) Tauc plots of $Zr_2CT_x$ (F) and $Hf_2CT_x$ (G) derived from UV-vis diffuse reflectance spectra, showing semiconducting nature of $Zr_2CT_x$ and $Hf_2CT_x$ with indirect bandgaps of 2.04 and 2.24 eV, respectively. Insets in (F) and (G) are the Kubelka-Munk transformed (F(R)) spectra as a function of wavelength.

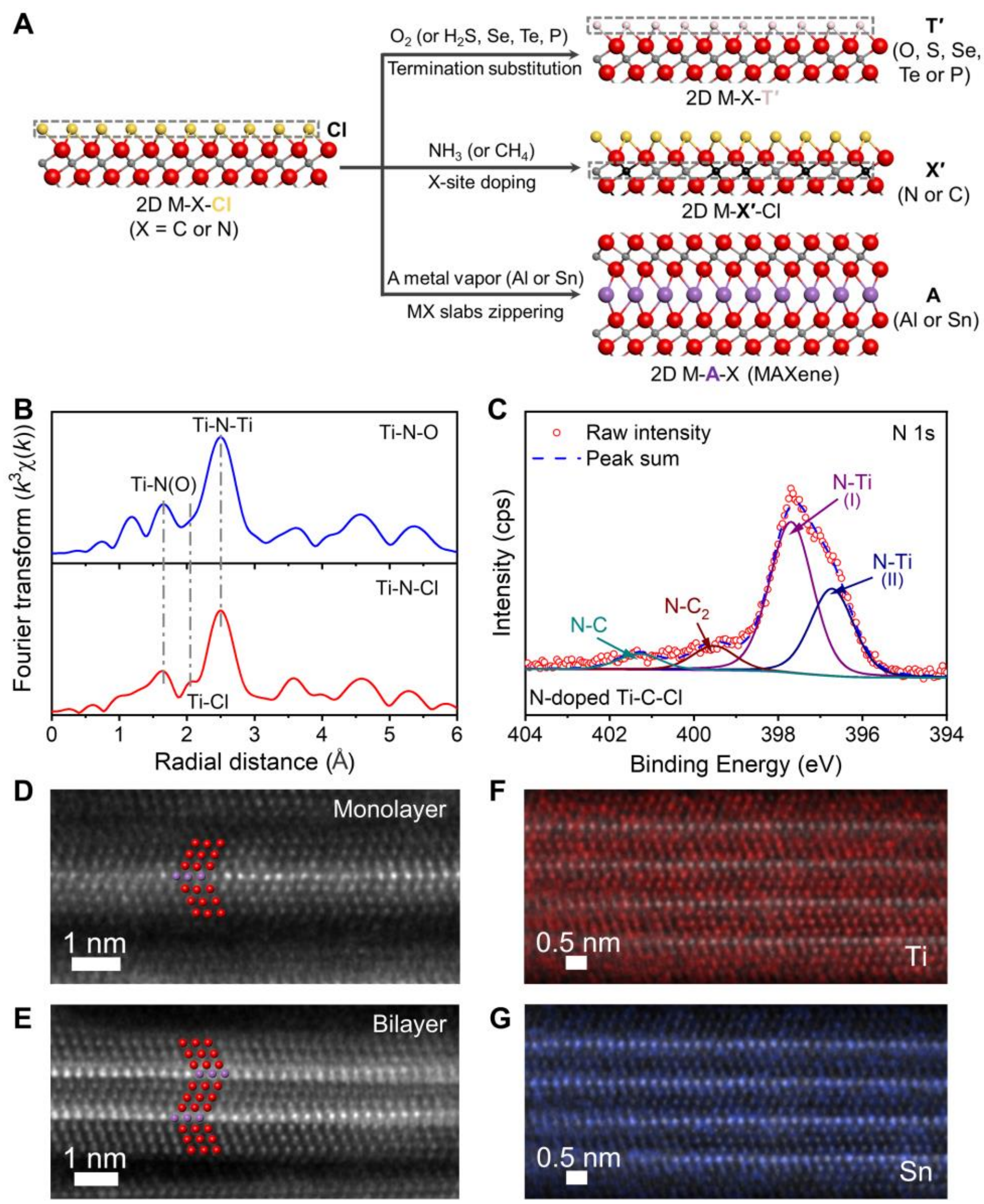

**Fig. 4. Compositional and structural tailoring of 2D layers via the vapor transformation strategy.** (**A**) Schematic illustration of the multi-dimensional tailoring of MXene layers with M-X-Cl configuration under available gases/vapors ($O_2$, $H_2S$, Se, Te, P, $NH_3$, $CH_4$, Al or Sn vapors), achieving termination substitution, X-site doping, and MX zippering. (**B**) Fourier transform spectra of Ti K-edge EXAFS spectra, exhibiting the disappearance of Ti-Cl peak and the increase of Ti-N/Ti-O peak for MXene with Ti-N-O configuration compared to Ti-N-Cl. (**C**) High-resolution N 1s XPS spectra of N-doped Ti-C-Cl, showing the characteristic peaks of N-Ti bonds after X-site doping. (**D** and **E**) Atomic-resolution STEM images of monolayer (D) and bilayer (E) of 2D atomic layers with Ti-Sn-C configuration. (**F** and **G**) Elemental mapping images of ultrathin layers with Ti-Sn-C configuration, showing the presence of Ti (F) and Sn (G) species.